\documentstyle[11pt,newpasp,twoside,epsf]{article}
\def\edcomment#1{\iffalse\marginpar{\raggedright\sl#1\/}\else\relax\fi}
\reversemarginpar

\begin{document}
\title{X-ray Properties of High Redshift Clusters}
\author{J. Patrick Henry}
\affil{Institute for Astronomy, University of Hawaii, 2680 Woodlawn Drive,
Honolulu, HI 96822, USA}

\begin{abstract}
We describe the ensemble X-ray properties of high redshift
clusters with emphasis on changes with respect to the local
population. Cluster X-ray luminosity evolution is detected in five
nearly independent surveys. The relevant issue now is characterizing
this evolution. Cluster temperature evolution provides constraints on
the dark matter and dark energy content of the universe. These
constraints are complementary to and in agreement with those of the
cosmic microwave background and supernovae, showing that the present
universe is dominated by a dark energy. X-ray images show that most
z $ > 0.75$ clusters are not relaxed, hinting that the cluster
formation epoch is z $ \sim 1$. 
\end{abstract}

\section{Introduction}

Twenty-five years ago almost all clusters of galaxies were selected
from surveys made at visible wavelengths. The two most widely used
catalogs were (and in many respects still are) those compiled by Abell
(1958) and Zwicky and Herzog (1983). A few dozen X-ray emitting
clusters were known then, almost all of them from the Abell catalog
and none at high redshift (here defined as z $ > 0.3$). 

This situation changed with the advent of focusing optics in X-ray
astronomy. The first X-ray detection of high redshift clusters was
with the Einstein Observatory (Henry et al., 1979), while the X-ray
selection of thousands of nearby clusters was made possible by the
ROSAT All-Sky Survey (see references in Section 1.3).

\subsection{Why Observe High Redshift Clusters?}

Because we want to measure the evolution of their properties. This
answer is obvious and begs the question why we want to do that.
Because the evolution of the cluster mass function is exponentially
sensitive to the values of several parameters of cosmological
interest. Further the evolution is simple, being driven by the gravity
of the underlying mass field of the universe and of a collisionless
collapse of cluster dark matter. We should be able to calculate this
evolution reliably. It is not possible to perform such calculations
for the only other objects visible at cosmological distances, galaxies
and AGNs.

A possible difficulty is that we usually do not observe mass. It
appears, however, that what we do observe in X-rays (see the next
section) has a good enough relation to mass that the difficulty can be
overcome. X-ray selection then offers two additional benefits.  There
is virtually no contamination from objects projected along the line of
sight and quantitiative selection criteria are easily established.

This fortunate situation of good theoretical understanding of a
relatively simple object coupled with the capability to make the
required observations has resulted in a great deal of work in this
field over the past several years. Examples include (Bahcall \& Fan,
1998; Blanchard et al., 2000; Donahue \& Voit, 1999; Henry, 2000;
Viana \& Liddle, 1999). All of this work used the venerable
Press-Schechter (Press \& Schechter, 1974 hereafter PS) mass function
on an open or spatially flat background universe. New theoretical
developments not yet fully incorporated are a more accurate
description of the mass function (Sheth \& Torman, 1999 hereafter ST)
and the generalization to an arbitrary cosmology (Pierpaoli, Scott, \&
White, 2001).

\subsection{X-ray Observables}
X-ray cluster observables are luminosity, temperature and surface
brightness (an image). The number of clusters for which these
observables has been obtained depends on the number of photons
required for a respectful measurement of them. A $5\sigma$ luminosity
measurment requires only about 25 photons. Hence many clusters have
their luminosity determined. A similar measurement of a cluster's mean
temperture needs $\sim 1000$ photons. Far fewer clusters have even an
average temperature measured, probably a few hundred, see White (2000)
for a large compilation. Finally, $\sim5000$ photons are required to
begin making a cluster image. Consequently there are very few high
redshift clusters with high statistics images.

In addition to observing individual objects, another complication
arises when measuring evolution. There are no standard clusters. There
is a tight correlation between X-ray isophotal size and temperature
(Mohr \& Evrard, 1997) and a loose correlation between luminosity and
temperature. Changes to these relations with redshift have been used
to search for evolution, but none have so far been found (Mohr et al.,
2000; Henry, 2000). The usual way to search for evolution is by
comparing the low and high redshift distributions of cluster
luminosities, the X-ray luminosity functions (XLFs), and cluster
temperatures, the X-ray temperature functions (XTFs).

\subsection{Samples of X-ray Selected Clusters}
We sumarize first the low redshift (z $ < 0.3$) local samples and then
the high redshift (z $ > 0.3$) samples.

The BCS consists of 201 clusters in the northern hemisphere detected
in the first processing (i.e. sorted into strips) of the ROSAT All-Sky
Survey (RASS) (Ebeling et al., 1998).  An extension to fainter fluxes,
the eBCS, has an additional 100 clusters (Ebeling et al., 2000). The
RASS1 Bright Sample contains 130 southern clusters, again from the
first processing of the RASS (De Grandi et al., 1999). REFLEX is
another southern hemisphere cluster survey, this time using the second
processing (i.e.  fully merged) RASS. It has 452 clusters (B\"ohringer
et al., 2001a).

For the high redshift samples we only give the number of objects with
z $ > 0.3$, although all but one of the following samples do have
objects at lower redshifts. The original serendipitous survey is the
EMSS (Gioia \& Luppino, 1994), the only survey described here coming
from the Einstein Observatory. The EMSS has 23 high redshift
clusters. Serendipitous in this context means that the clusters were
found in the fields of observations targeted at other objects. Most
high redshift X-ray selected clusters are found in this way as it is
the only way to go deep enough. A major drawback of this approach is
the solid angle surveyed is not contiguous. All but two of the
following surveys use this strategy with the ROSAT pointed data.

The 160 Square Degree Survey has 73 z $ > 0.3$ clusters (Vikhlinin et
al., 1998). The original catalog gave spectroscopic redshifts for 36\%
of the clusters, but we have since obtained them for all but one
object (Mullis et al., 2002). With 12 objects at high redshift, the
BRIGHT SHARC has the smallest sample size (Romer et al., 2000). The
RDCS contains 50 clusters at z $> 0.3$ to the lowest flux limit of all
the surveys (Borgani et al., 2001). WARPS contains $\sim 75$ high
redshift objects (Jones et al., 2000).

There are two high redshift cluster surveys that use the RASS and
hence are the only ones that have contiguous sky coverage. The NEP
survey covers the deepest region of the RASS around the North Ecliptic
Pole (Henry et al., 2001).  It contains 19 high redshift clusters.
MACS surveys 55\% of the sky for the most luminous clusters at
z $ > 0.3$. It has 34 objects in its bright subsample and $ >75$ additional
clusters above a flux limit a factor of 2 lower (Ebeling, Edge \& Henry,
2001). Already MACS is the largest single X-ray selected sample of
clusters with z $> 0.3$. 

The last survey we describe has just begun. The BMW uses data from the
ROSAT HRI pointings, a previously neglected archive (Moretti et al.,
2001). The distinct advantage of this survey is the HRI has much
better spatial resolution compared to the PSPC, which was used for all
ROSAT cluster surveys until now. Clusters are easily selected by X-ray
extent with the HRI. Somewhat surprisingly, given the high HRI
background, the flux limit of the BMW is competitive with other
surveys.

\begin{figure}
\plottwo{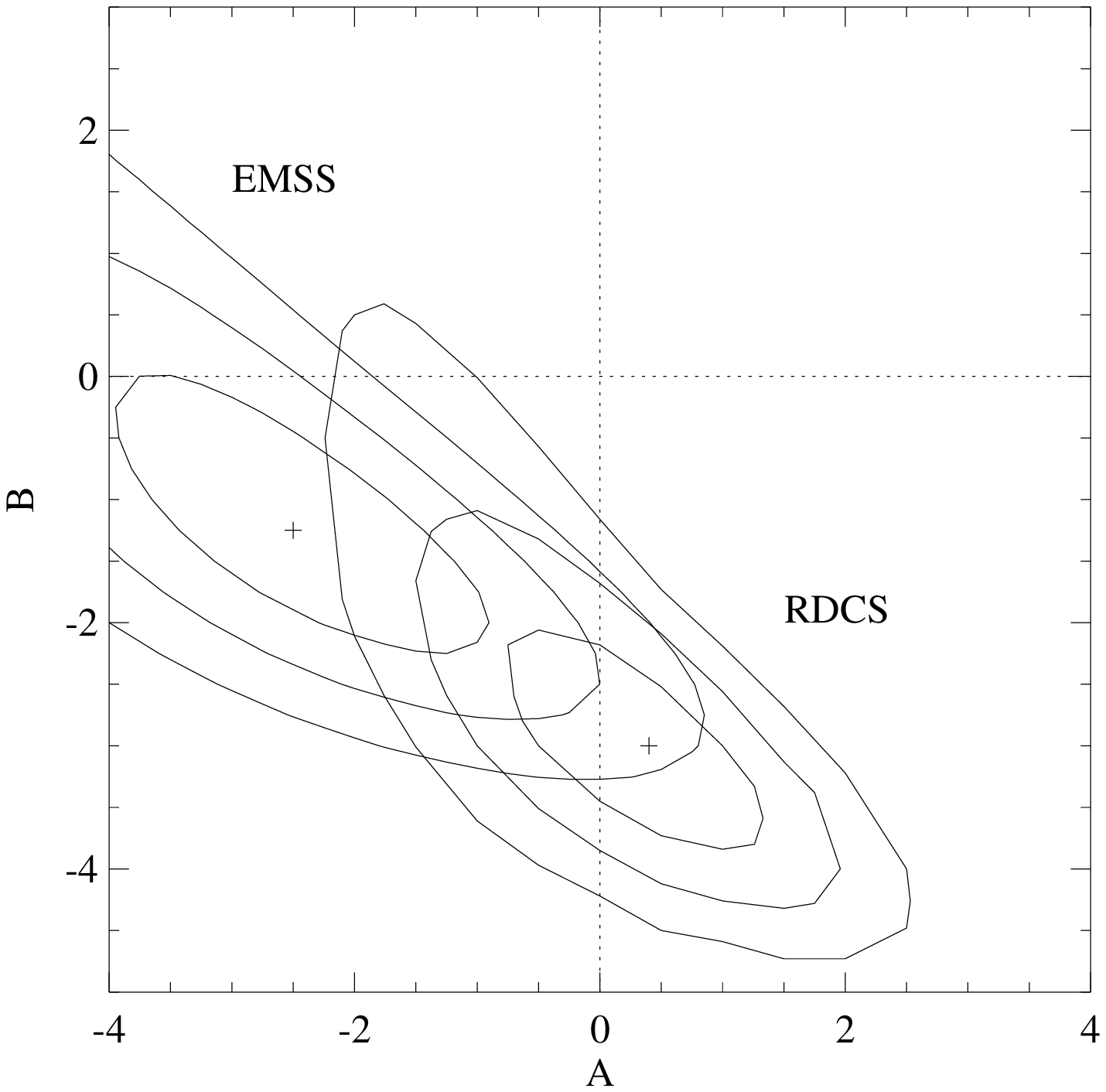}{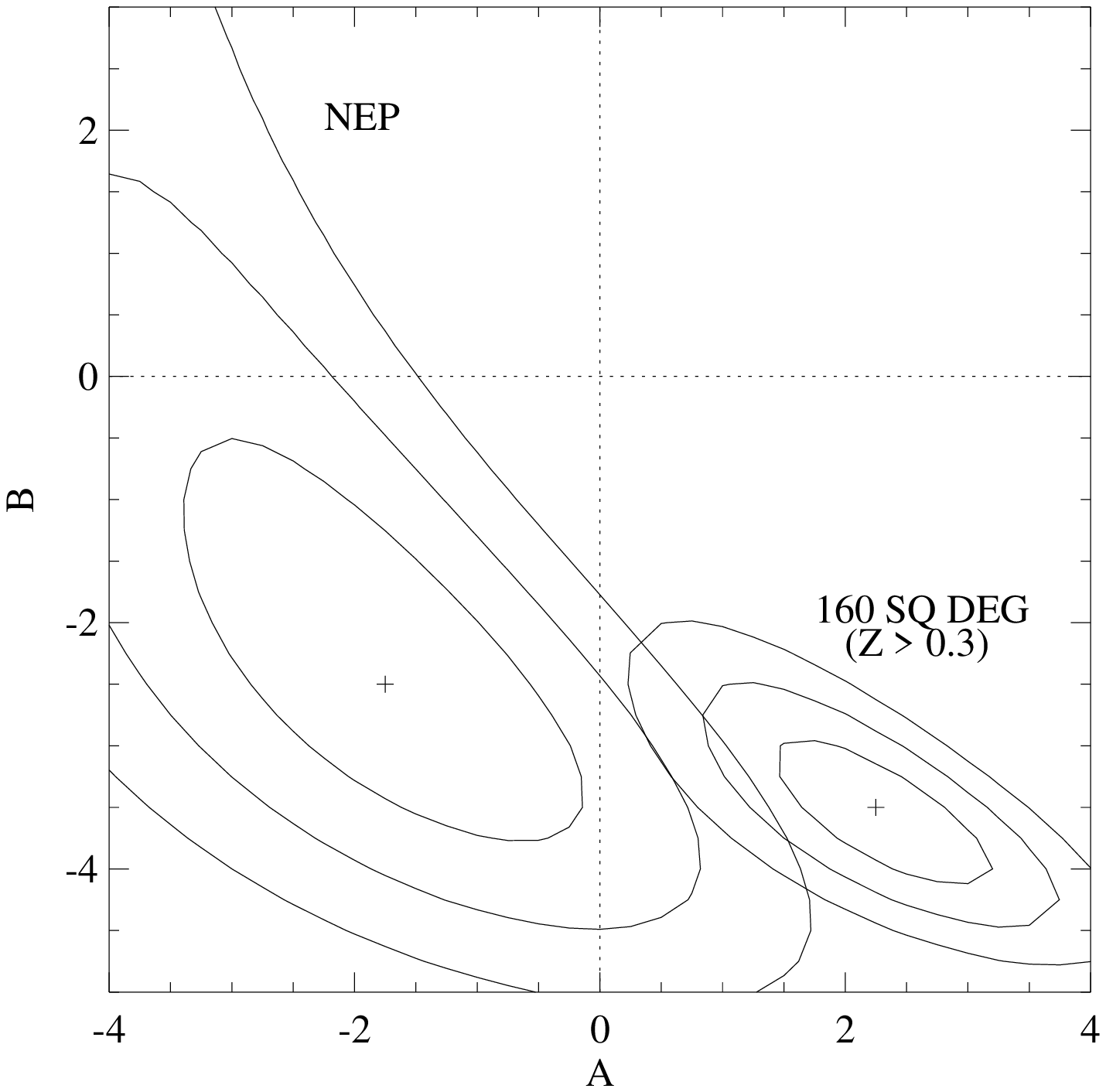}
\caption{a. (left) One, two and three $\sigma$ contours of the AB
model fit to the EMSS and RDCS surveys. b. (right) One, two and three
$\sigma$ contours of the AB model fit to the 160 deg$^{2}$ z $ > 0.3$
and NEP surveys.}
\end{figure}

\section{Cluster X-ray Luminosity Evolution}

Evidence for evolution of the luminosities of clusters of galaxies
came originally from the EMSS (Gioia et al. 1990; Henry et
al. 1992). These studies found that the co-moving number density of
high luminosity clusters is smaller in the past than at
present. However, the result was barely significant and it has enjoyed
healthy skeptism. Many more samples are now available to confirm or
refute this claim.

\subsection{Six High Redshift Samples Analyzed the Same Way}
Each of the surveys described in Section 1.3 has a unique selection
function that must be removed in order to compare them. The usual
method, ploting luminosity functions, does not use all the information
available. Since the evolution seems to be a lack of objects at high
redshifts, there is nothing to plot if the objects are not
there. Instead we perform maximum likelihood fits of five high
redshift samples to the AB model introduced by Rosati et al. (2000)
who has already analyzed the RDCS.  In this model the XLF is an
evolving Schechter function: $n(L,z) = n_{0}(z) L^{-\alpha}
e^{-L/L^{*}(z)}$, with $n_{0}(z) = n_{0} [(1+z)/(1+z_{0})]^{A}$ and
$L^{*}(z) = L^{*}_{0} [(1+z)/(1+z_{0})]^{B}$. Note that $n_{0},
\alpha$ and $L^{*}_{0}$ are not fit, but come from a low redshift XLF,
in this case the BCS since it has the lowest normalization of the
three local determinations thus yielding the least evolution. We set
$z_{0}$ to 0.1, the characteristic redshift of the BCS. No evolution
in this model is the point $A = B = 0$. Note further that a maximum
likelihood fit incorporates the information provided by any
``missing'' high redshift clusters. We assume, to be consistent with
previous work, that $H_{0} = 50$ km s$^{-1}$ Mpc$^{-1}$ and $q_0 =
0.5$, where $H_0$ is the the Hubble parameter and $q_0$ is the
deceleration parameter, both at the present epoch.

\begin{figure}
\plottwo{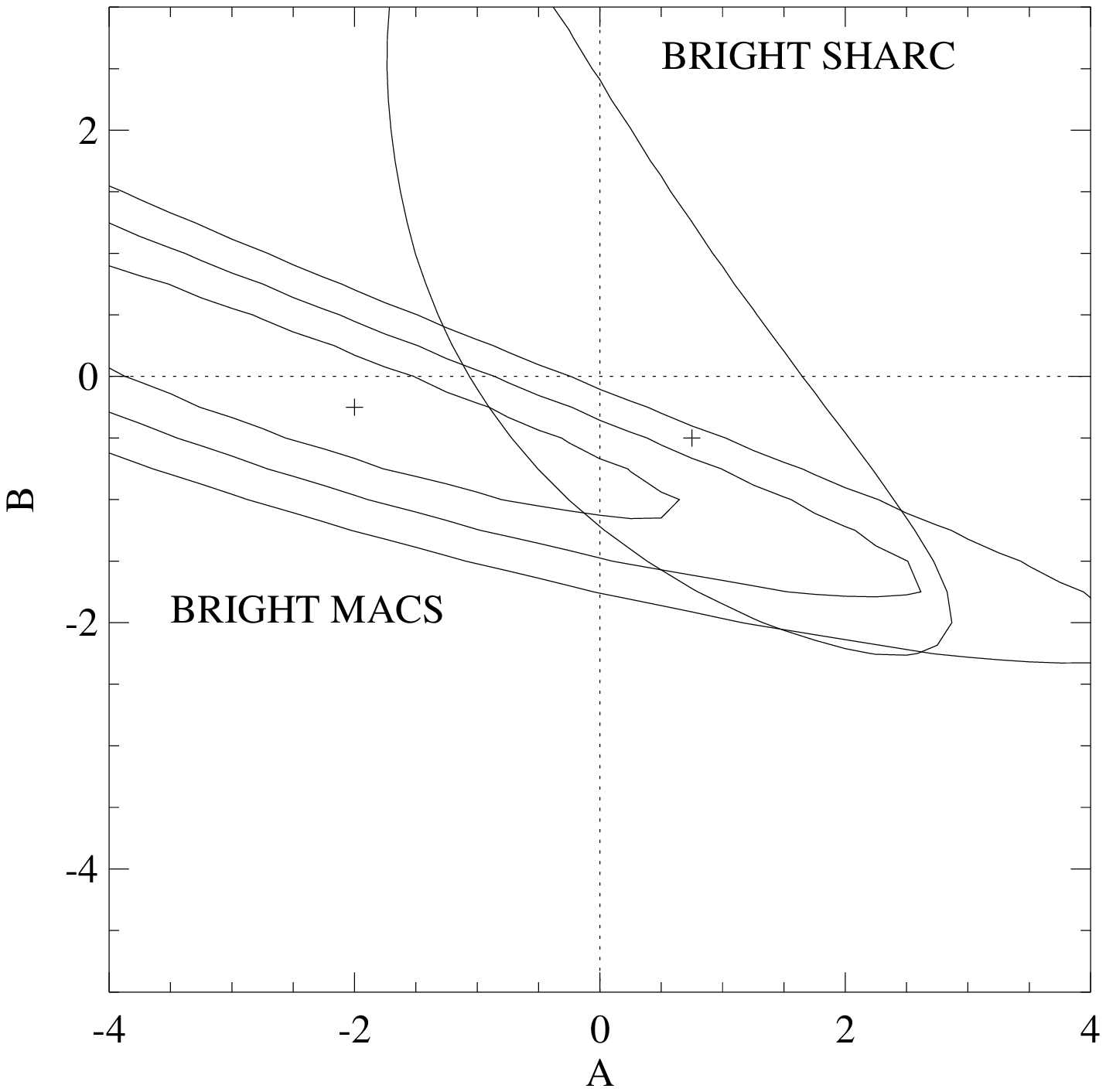}{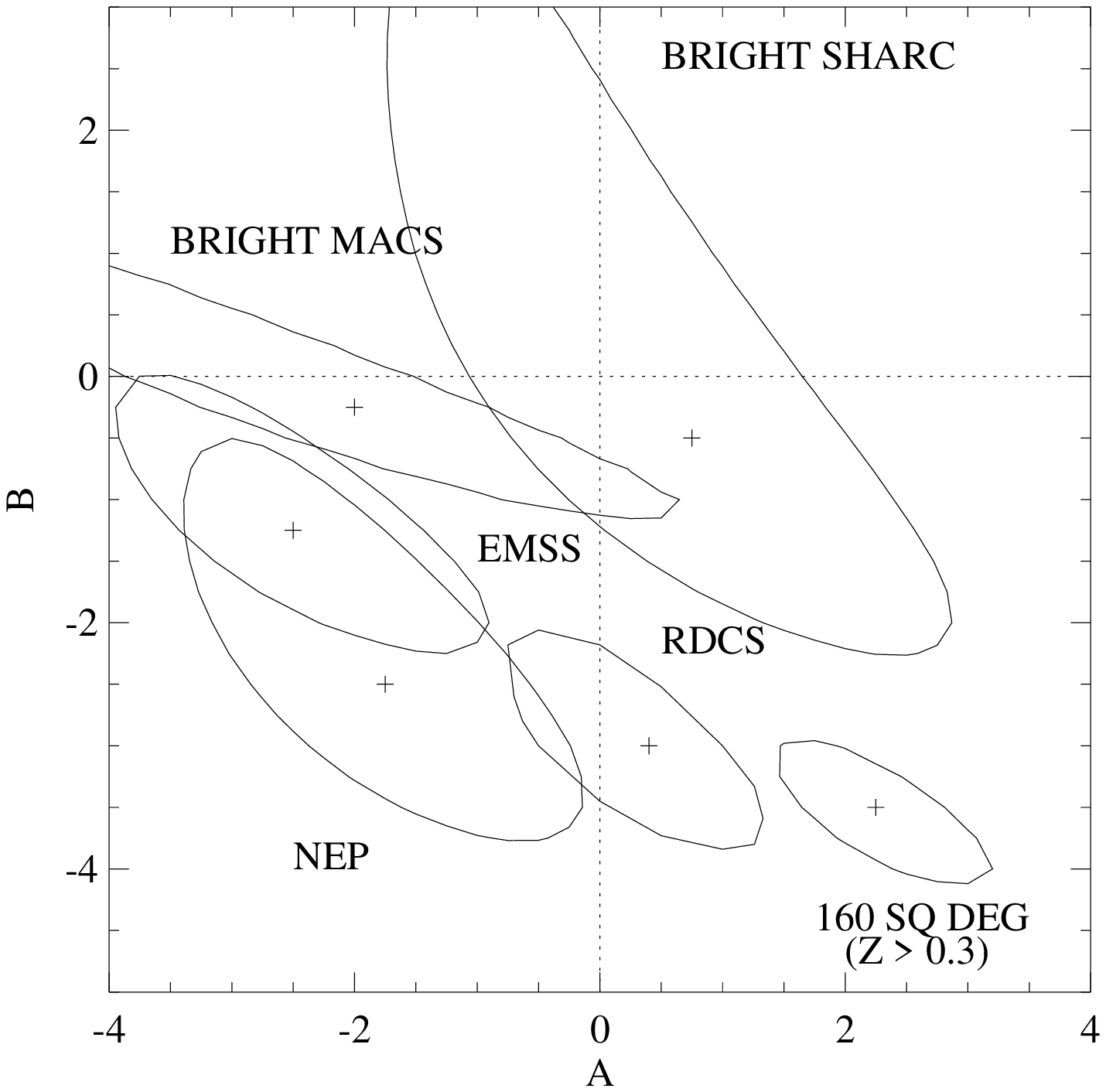}
\caption{a. (left) One, two and three $\sigma$ contours of the AB
model fit to the Bright MACS and one $\sigma$ contour for the Bright
SHARC surveys. b. (right) One $\sigma$ contours of the AB model fit to
the EMSS, RDCS, NEP, Bright MACS, Bright SHARC and 160 deg$^{2}$ z $>
0.3$ surveys.}
\end{figure}

We show the results of the fits in Figures 1, and 2. Five of the six
samples exhibit luminosity evolution at the $> 3 \sigma$ level. The
only survey that does not exhibit significant evidence for evolution
is the Bright SHARC, which is the smallest sample. Figure 2b
shows that some of the surveys agree at the $\sim 1 \sigma$ level,
e.g.  Bright MACS, EMSS and NEP and RDCS, EMSS and NEP. However, on
the whole the agreement among all the surveys is marginal at
best. More work will be required to determine whether this diagreement
is real and/or results from the specific model fitted. In particular,
we have forced the best fitting low redshift XLF onto the fit without
considering the errors in its parameters. Further, the model probably
is too simple since it assumes that all clusters in the low redshift
sample are at the same redshift, $z_{0}$. In reality, because all
samples are flux limited, different redshift clusters determine the
different Schechter function fit parameters. See also the end of
Section 2.2 for additional discussion of this point.

\begin{figure}
\plotfiddle{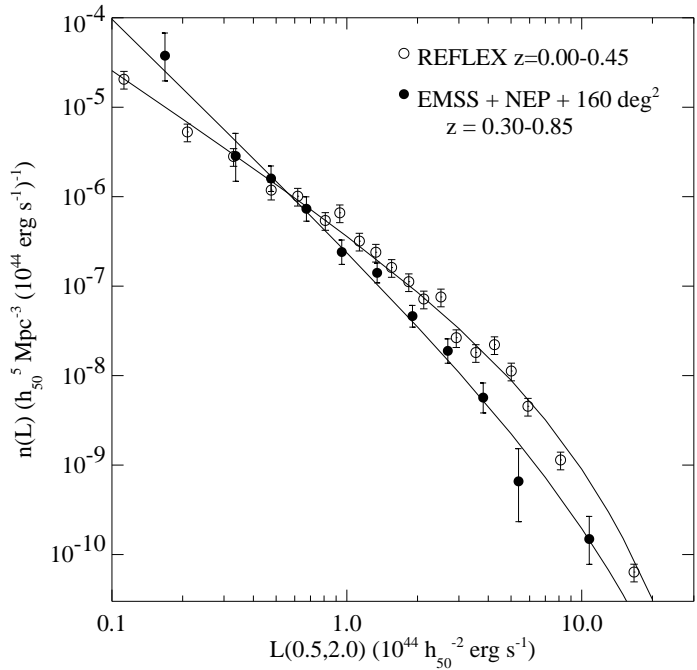}{3.3in}{0}{60}{60}{-140}{-10}
\caption{Comparison of the low z XLF from the REFLEX survey with the
high z XLF from the combined EMSS, NEP and 160 deg$^{2}$ surveys. The
best fitting Schechter function XLFs are overlaid}
\end{figure}

\subsection{Combining Three High Redshift Samples for Better Statistics}
The fits described in Section 2.1 show that there is no longer a
question whether cluster X-ray luminosities evolve, the five surveys
with best statistics show they do. Rather the question now is how to
characterize that eveolution.

We therefore construct the high redshift cluster luminosity function
from the sum of the EMSS (plus RX J0152.7-1357), NEP, and 160
deg$^{2}$ samples in order to obtain a higher statistics nonparametric
description of that evolution. There are 110 objects in this combined
sample, comparable at last to the low redshift samples. The average
redshift of clusters in the combined sample is 0.45. The overlap on
the sky of these three samples is about 5\%, so we have corrected
statistically for double counting since the corrections are not
large. We compare in Figure 3 this high z XLF to the REFLEX low z XLF
from B\"ohringer et al. (2001). We use REFLEX for this comparison
because it is the largest local sample. The raw luminosity functions
clearly show the negative evolution at high luminosities. There may be
positive evolution at low luminosities, but only one low statistic
luminosity bin exhibits this effect. The high z XLF falls a factor of
$\sim 2.5$ below the REFLEX XLF at a luminosity of $2 \times 10^{44}$
erg s$^{-1}$ in the 0.5 - 2.0 keV band.

\begin{figure}
\plotfiddle{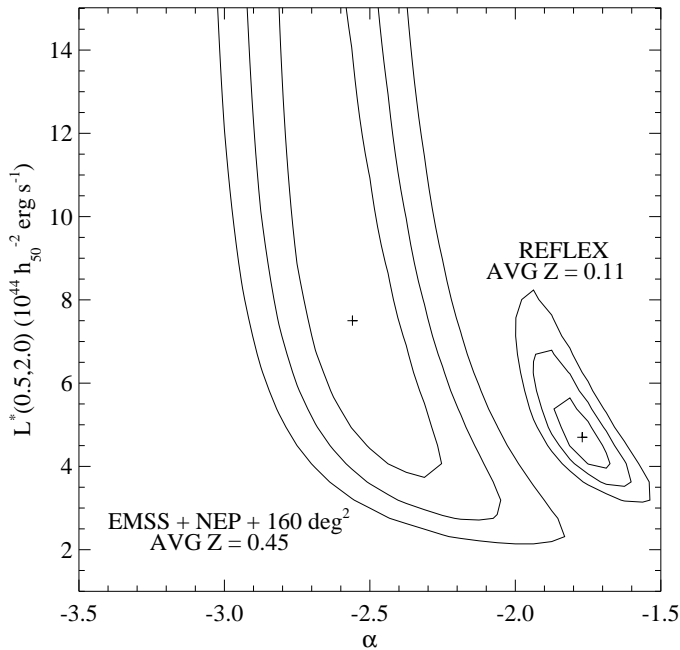}{3.3in}{0}{60}{60}{-140}{-10}
\caption{One, two and three $\sigma$ contours for the two interesting
parameters of Schechter model fits to the REFLEX and combined EMSS,
NEP and 160 deg$^{2}$ surveys. The small ($\sim 1 \sigma$) difference
between our best fit for the REFLEX sample and that of B\"ohringer et
al. (2001b) comes from our restricting the fit to clusters with
luminosities in the 0.1 to 2.4 keV band above $1 \times 10^{43}$
$\mathrm {erg \:s^{-1}}$ and redshifts below 0.3.}
\end{figure}

As a first attempt at characterizing the observed evolution, we have
performed maximum likelihood fits of Schechter functions to the two
samples. We show the best fits in Figure 3. Figure 4 gives the
likelihood contours in the $L^{*}$ - $\alpha$ plane. The best fitting
normalization, $n_{0}$, is that which yields the observed number of
clusters in the sample when integrating the Schechter function with
best fitting $L^{*}$ and $\alpha$ over the survey selection
function. This analysis demonstrates that at least $n_{0}$, and
$\alpha$ are evolving.  $L^{*}$ may be evolving, but is not required
to be. Recall that the AB model has $n_{0}$ and $L^{*}$ evolving.

\section{Constraining Cosmology From Cluster Temperature Evolution}

A cluster's luminosity is proportional to the square of its X-ray gas
density. Hence luminosity evolution is very sensitive to the evolution
of the gas, which is almost certainly different from the evolution
driven by gravity during hierarchical growth. Temperature is probably
the X-ray observable most closely related to mass and should more
accurately reflect this growth.

\begin{figure}
\plotfiddle{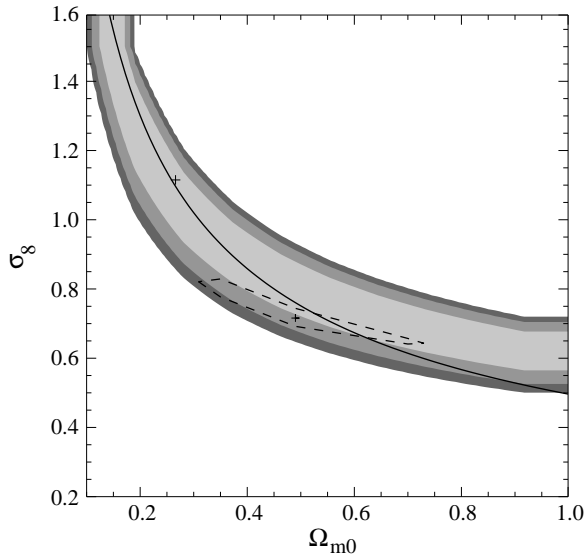}{2.7in}{0}{60}{60}{-190}{-130}
\caption{Constraints in the $\sigma_8$ - $\Omega_{m0}$ plane for an
open universe. The shaded region gives the one, two and three $\sigma$
contours from an analysis of weak lensing by field galaxies (Maoli
et al. 2001). The cross is their best fitting value. The line is from
an analysis of the local cluster temperature function by Pierpaoli et
al. (2001). The dashed ellipse and cross is from an analysis of
cluster temperature evolution by Henry (2000). There is good agreement
among the three results. Original figure from Maoli et al. (2001)}
\end{figure}

There are two cosmological parameters that mainly determine the
hierarchical growth and thus can be strongly constrained by cluster
temperature evolution. These two are $\Omega_{m0}$, the present mass
density relative to the critical density and $\sigma_8$, the amplitude
of mass density fluctuations on a scale of $8 h^{-1}$ Mpc. This latter
parameter is a complicated way to normalize the present value of the
power spectrum of mass density fluctuations, P(k), at $k \approx 0.2 h
Mpc^{-1}$: $\sigma_8 \approx [P(0.172 h Mpc^{-1})/3879 h^{-3}
Mpc^3]^{1/2}$ (Peacock, 1999 equations 16.13 and 16.132). Note that
the rate of hierarchical growth is not very dependent on
$\Omega_{\Lambda0} = 3\Lambda/H_0^2$ with $\Lambda$ the cosmological
constant. Cluster evolution does not constrain this parameter very
well.

\subsection{Press-Schechter Mass Function in a Flat Universe}
Many papers over the last several years have analyzed cluster
temperatures with the aim of constraining cosmology. We reference a
selection of them in Section 1.1. To date all of this work has assumed
a background cosmology that is either flat ($\Omega_{m0} +
\Omega_{\Lambda0} = 1$) or open ($\Omega_{\Lambda0} = 0$).  The mass
function is given by the PS formula. Mass is converted to temperature
by assuming clusters form via a top hat collapse: $kT = 1.42
[\Omega_{m0} \Delta(\Omega_{m0},z)]^{1/3} [hM_{15}]^{2/3} [1
+z]/\beta$.  Here $\Delta(\Omega_{m0},z)$ is the ratio of the
cluster's average density to that of the background mass density at
the virialization redshift, $M_{15}$ is the virial mass of the cluster
in units of $10^{15}M_{\sun}$ and $\beta$ is a modification factor
accounting for departures from hydrostatic equilibrium. It is
determined from numerical hydrodynamic simulations. Our work uses
$\beta = 1.21$.

\begin{figure}
\plotfiddle{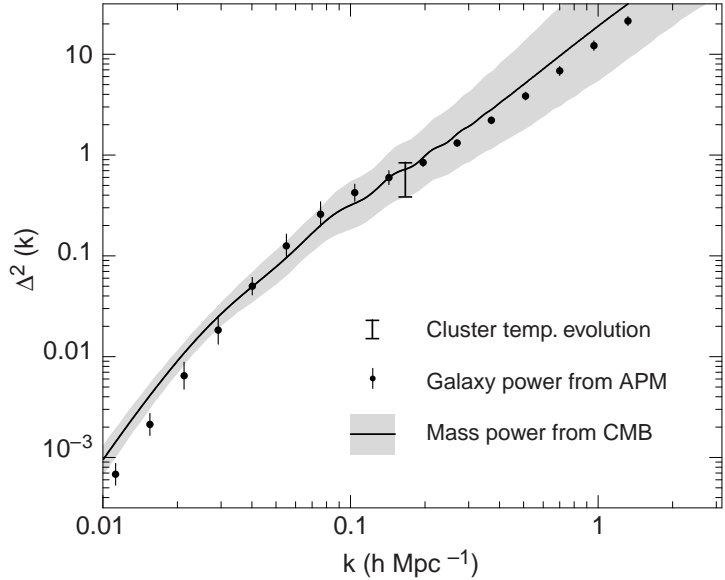}{2.7in}{0}{60}{60}{-190}{-140}
\caption{Fluctuation power spectra where
$\Delta^2(k)=V/(2\pi^2)k^3P(k)$, but $\sigma_8 \simeq
\Delta(0.172hMpc^{-1})$.  The shaded region gives the one $\sigma$
error on the present mass fluctuation power spectrum predicted by
evolving the measured microwave background fluctuation spectrum to the
present, under the assumption that $H_0 = 70$ km s$^{-1}$ Mpc$^{-1}$
$\pm 10\%$. The dots with errors are the galaxy fluctuation spectrum,
while the single error bar is the value of $\sigma_8$ from an analysis
of cluster temperature evolution by Henry (2000). The good agreement
between the predicted spectrum and its observed normalization implies
that clusters form from hierarchical growth. Original figure from
Peacock et al. (2001)}
\end{figure}

The local abundance of clusters yields the following constraint:
$\sigma_8 \Omega_{m0}^{0.5} \simeq 0.5$. Figure 5 shows a recent
example. Measuring the abundance at two epochs, that is measuring
evolution, breaks this degeneracy and is a prime reason for making
such measurements.

As an example, we give the results of our recent work (Henry 2000),
which is fairly typical. None of the local samples described in
Section 1.3 have complete temperature information, so we used a sample
of 25 clusters originally from Piccinotti et al. (1982) with
corrections for source confusion. Sixteen of these objects had
temperatures from ASCA when we did the analysis. The only high
redshift sample that has complete temperature information is the
EMSS. We used 14 EMSS clusters selected by flux, all with ASCA
temperatures and ROSAT HRI images. The average redshift of the local
sample is 0.05 while that of the distant sample is 0.38. We found that
$\Omega_{m0} = 0.49 \pm 0.12 (1\sigma) \pm 0.23 (2\sigma)$ and
$\Omega_{m0} = 0.44 \pm 0.12 (1\sigma) \pm 0.23 (2\sigma)$ for open
and flat cosmologies respectively. $\Omega_{m0} = 1$ is excluded at
$>> 99\%$ confidence. For $\sigma_8$ we find $0.72 \pm 0.10$ and $0.77
\pm 0.15 (1 \sigma)$ for open and flat models respectively. 

We show in Figures 5 and 6 a comparison of these results with some
recent work not available when Henry (2000) was completed. Figure 5
compares the results deduced form weak gravitational lensing by field
galaxies (Maoli et al., 2001) with those obtained from the local
abundance of clusters (Pierpaoli et al., 2001). There is very good
agreement with a similar degeneracy. Our results break the degeneracy
as described above, again with very good agreement. Figure 6 shows the
present power spectrum of density fluctuations (Peacock et al.,
2001). The mass power spectrum is that predicted from cosmic microwave
background, i.e. evolved from a redshift of $\sim1000$ to the present,
plus the assumption that $H_0 = 70$ km s$^{-1}$ Mpc$^{-1}$ $\pm
10\%$. The galaxy spectrum shows that galaxies exhibit a small bias
with respect to the mass. The agreement between the cluster
temperature normalization and the predicted mass spectrum shows that
the tiny fluctuations present in the universe at $z \sim 1000$ really
have grown by gravity to produce the clusters we observe today.

\begin{figure}
\plotfiddle{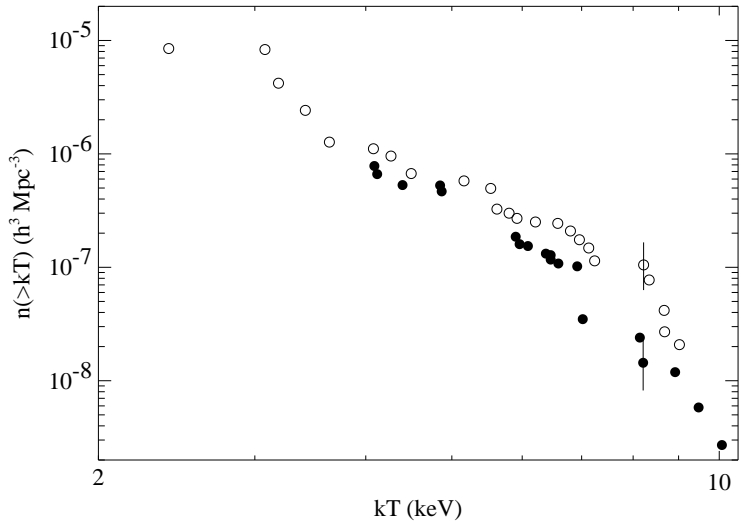}{2.6in}{0}{60}{60}{-150}{-10}
\caption{Integral cluster temperature function. The open (filled)
circles are the low (high) redshift cluster observations. The high
redshift data come from the entire EMSS z $ > 0.3$ sample plus RX
J0152.7-1357.}
\end{figure}

\subsection{Improved Mass Function with Arbitrary $\Omega_{m0}$ and
$\Omega_{\Lambda0}$} 

Although the PS mass function is a good approximation to that obtained
from large N-body simulations, it does predict too many low-mass
clusters and too few high mass ones (see Jenkins et al., 2001). The ST
mass function does a better job although it is not quite as good as
the Jenkins et al.  empirical fit. The ST mass function allows
clusters to form via an ellipsoidal rather than the spherical PS
collapse. We prefer the ST function because the empirical fits will
presumably change with different numerical simulations; ST has the
virtue that it stays the same.  Another advance is the function
$\Delta$ in the mass - temperature relation has finally been
determined for arbitrary $\Omega_{m0}$ and $\Omega_{\Lambda0}$
(Pierpaoli et al., 2001).

We have made a new determination of cluster temperature evolution. The
local sample remains the same 25 clusters used in Henry (2000).  All
their temperatures are measured with ASCA by White (2000), thereby
eliminating a possible systematic effect since all temperatures (but
one in the high redshift sample) are measured with the same
experiment. The high redshift sample is now all EMSS clusters with z
$> 0.3$ plus RX J0152.7-1357 with a temperature from Della Ceca et
al. (2000). There are 20 clusters in the distant sample, the average
redshift of which has increased to 0.45. We show the temperature
functions in Figure 7. The temperatures exhibit the same moderate
evolution as the luminosities discussed in Section 2.

\begin{figure}
\plotfiddle{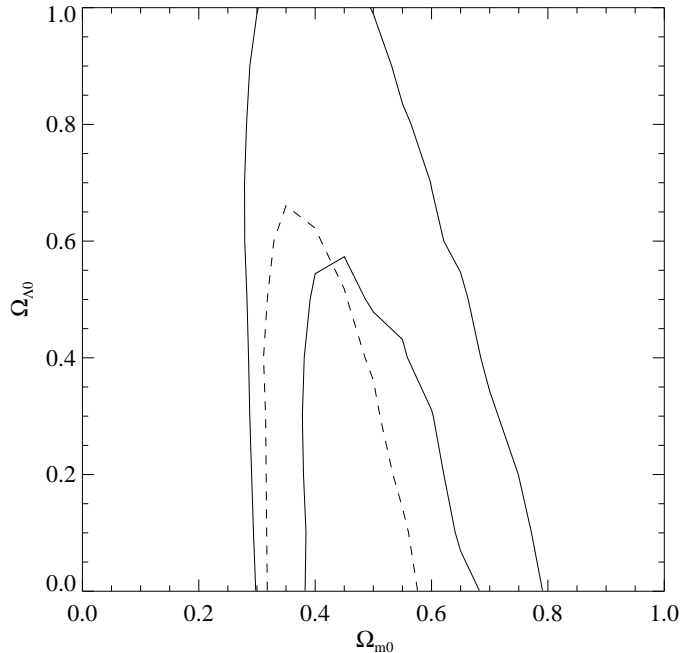}{3.2in}{0}{60}{60}{-150}{-10}
\caption{Constraints on the dark energy and dark matter content of the
universe provided by the data in Figure 7. The solid lines are the one
and two $\sigma$ contours using the ST mass function, while the dashed
line is the one $\sigma$ contour from the PS mass function. The
difference is small, $\sim 0.5 \sigma$.}
\end{figure}

The recent theoretical advances permit us to constrain the geometry of
the universe. We show these constraints in Figures 8 and 9. Figure
8 shows that the difference between the PS and ST mass functions is
minor for the present sample size, $\sim 0.5 \sigma$. Figure 9
compares the constraints provided by supernovae, the cosmic microwave
background and cluster evolution. All three methods are concordant and
nicely complementary, thereby diminishing the possibility that any one
of them has significant systematic errors. It is clear that 
$\Omega_{m0} \sim 0.35$ and $\Omega_{\Lambda0} \sim 0.65$ agrees with
all determinations. It appears that the universe has significant amounts
of dark matter and dark energy at present.

\begin{figure}
\plotfiddle{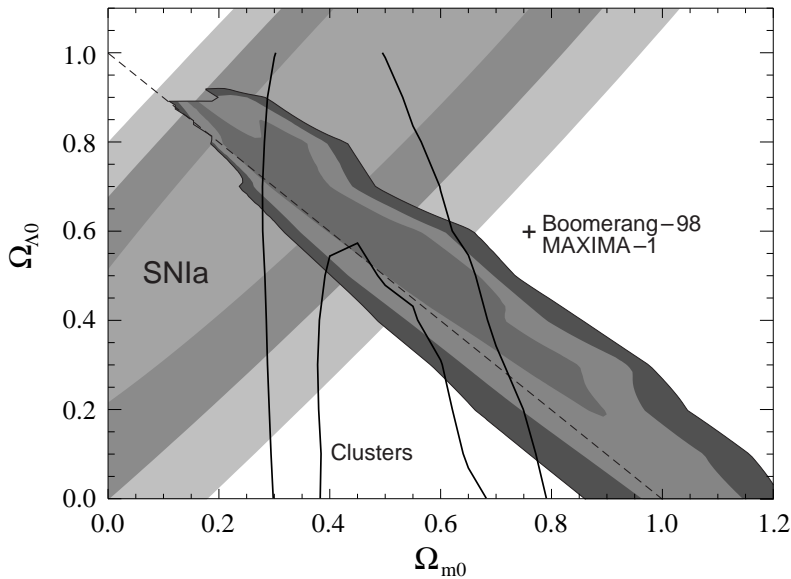}{2.9in}{0}{80}{80}{-250}{-290}
\caption{Comparison of the constraints provided by cluster temperature
evolution, supernovae and the microwave background. All three
provide complementary constraints that show that the universe is
presently dominated by dark energy. Original figure from Jaffe et al.
(2001).}
\end{figure}

\subsection{Summary of Cluster Temperature Evolution Constraints of
Cosmological Parameters} Although the agreement exhibited by Figures
5, 6, and 9 is encouraging, there is a rather wider range in the best
fit value of $\Omega_{m0}$ found by different investigators than would
be expected from the quoted errors.  One uncertainity is that clusters
are selected by their flux, not temperature. The selection is
quantified by specifying the surveyed solid angle as a function of
flux. The flux can be straightforwardly converted to luminosity and
redshift. The luminosity may be converted to temperature from the
empirical cluster luminosity - temperature relation yielding the solid
angle as a function of temperature and redshift.  This second step is
not as straightforward as the first. Although there is a relation, it
has large scatter. At least the relation does not seem to evolve very
much. The community has not yet settled on a preferred analysis method
that incorporates this uncertainity.  There is good agreement on the
value of $\sigma_8$ among the various investigators or at least most
results fall along the same $\sigma_8$ - $\Omega_{m0}$ degeneracy curve.

\section{State of Very High Redshift Clusters}

There are now about a dozen X-ray emitting clusters presently known at
z $> 0.75$. Chandra images are in hand for 8 of them (e.g. Cagnoni et
al., 2001; Fabian et al., 2001; Jeltema et al., 2001; Stanford et al.,
2001).  All but 2 of these 8 are obviously not relaxed. This situation
contrasts with that at low redshift where most clusters appear to be
much closer to equilibrium (see the compilation of images in Mohr,
Mathiesen, \& Evrard, 1999). Although qualitative, this striking
difference leads us to suspect that z $\sim 1$ is more likely the
epoch of cluster formation than is the present.

\acknowledgements Thanks are due to the organizers of our conference for
a great meeting in the magnificant setting of the Taroko National
Park. I also want to thank my many collaborators with whom I
have worked on cluster surveys over the years. These include:
I. Gioia, C. Mullis, W. Voges, U. Briel, H. B\"ohringer and J. Huchra
for the NEP; A. Vikhlinin, C. Mullis, I. Gioia, B. McNamara,
A. Hornstrup, H. Quintana, K. Whitman, W. Forman, and C. Jones for the
160 deg$^{2}$; H. Ebeling and A. Edge for the MACS. Most of the work
described here will eventually appear as publications coauthored with
them. H. B\"ohringer and H. Ebeling provided the digital data from 
Figures 20 and 10 of B\"ohringer et al. (2001a) and Ebeling et al.
(2001) respectively.

\end{document}